\documentclass[aps,prb,twocolumn,showpacs,10pt,superscriptaddress]{revtex4-1}
\usepackage{graphicx,amsmath,amssymb,amsfonts,sidecap,color}

\newcommand{\ket}[1]{\ensuremath{\left|#1\right\rangle}}

\makeatletter
\newcommand*{\rom}[1]{\expandafter\@slowromancap\romannumeral #1@}
\makeatother

\begin{document}
 
 \title{Kramers Pairs of Majorana Fermions and Parafermions \\
in  Fractional Topological Insulators}

\author{Jelena Klinovaja}
\affiliation{Department of Physics, Harvard University,  Cambridge, Massachusetts 02138, USA}
\author{Amir Yacoby}
\affiliation{Department of Physics, Harvard University,  Cambridge, Massachusetts 02138, USA}
\author{Daniel Loss}
\affiliation{Department of Physics, University of Basel,
            Klingelbergstrasse 82, CH-4056 Basel, Switzerland}

\date{\today}
\pacs{71.10.Pm; 74.45.+c; 05.30.Pr}

\begin{abstract}
We propose a scheme based on topological insulators to generate Kramers pairs of Majorana fermions or parafermions in the complete absence of magnetic fields. Our setup consists of two topological insulators whose edge states are brought close to an $s$-wave superconductor.  The resulting proximity effect leads to an interplay between a non-local crossed Andreev pairing, which is dominant in the strong electron-electron interaction regime, and usual superconducting pairing, which is dominant at large separation between the two topological insulator edges.  As a result, there are zero-energy bound states localized at interfaces between spatial regions dominated by the two different types of pairing. Due to the preserved time-reversal symmetry, the bound states come in Kramers pairs.
If the topological insulators carry fractional edge states, the zero-energy bound states are parafermions, otherwise, they are Majorana fermions.

\end{abstract}

\maketitle

\section{Introduction}

During the last decade  we have witnessed a rapidly growing interest in  quantum states in lower dimensions with non-trivial braiding statistics. In particular bound states at zero-energy have attracted a lot of attention. For instance, a plethora of new setups that can host Majorana fermions (MFs), particles that are their own antiparticles, has been suggested
\cite{Read_2000,fu,Nagaosa_2009,Sato,demler_2011,potter_majoranas_2011,
lutchyn_majorana_wire_2010,oreg_majorana_wire_2010,Rotating_field,alicea_majoranas_2010,
RKKY_Basel,RKKY_Simon,RKKY_Franz,Klinovaja_CNT,bilayer_MF_2012,MF_nanoribbon,MF_MOS,MF_Bena,MF_ee_Suhas} and motivated many experiments to
test these predictions.\cite{Ando,mourik_signatures_2012,deng_observation_2012,das_evidence_2012,Rokhinson,Goldhaber,marcus_MF}
These exotic particles are interesting not only from a fundamental point of view  but also find a direct application 
in topological quantum computing schemes.\cite{Alicea_braiding}  Majorana fermions are non-Abelian anyons with a braiding statistics of the Ising-type. Such anyons allow to implement some of fundamental quantum operations but not all of them.\cite{Pachos_book}  To extend the class of operations, one can work with parafermions (PFs) that possess even more exotic braiding statistics arising in the presence of strong electron-electron interactions.\cite{Fradkin_PF_1980,topology_barkeshli,Fendley_PF_2012,PF_Linder,Cheng,Vaezi,PF_Clarke,
Ady_FMF,PF_Mong,vaezi_2,PFs_Loss,PFs_Loss_2} So far, the search for most experimentally feasible setups continues as well as a discussion on which path to follow. On one hand, one can generate MFs and try to supplement their braiding statistics with non-topological gates. On the other hand, one can aim at generating non-Abelian braiding statistics of the Fibonacci type from a sea of parafermions.\cite{PF_Mong}

In the present work we focus on Kramers pairs of Majorana fermions\cite{TRI1,TRI2,TRI3,TRI4,TRI5,TRI6,TRI7,TRI8,TRI9,TRI10,PFs_Loss_2} and parafermions.\cite{PFs_Loss_2} The basic ingredients of our setup are two topological insulators\cite{Hasan_review,Volkov_TI1,Volkov_TI2,Volkov_TI3,Fu_Kane,Zhang_TI,exp_1,Konig_2 , Konig, Roth_TI,Nowack_TI,Amir_TI,Patric_TI,Carlos_TI} brought into proximity to a superconductor. We assume that the chemical potential lies in the topological gap, so we focus only on edge modes, signatures of which were recently observed in experiments.\cite{exp_1,Konig_2 , Konig, Roth_TI,Nowack_TI,Amir_TI} We also note that the proximity coupling between superconductors and topological insulators were also successfully implemented.\cite{Amir_TI}
Schemes for Majorana fermions in topological insulators with proximity effect have been first proposed by Fu and Kane\cite{fu} and implemented in recent experiments\cite{Ando,Goldhaber}, where a finite magnetic field was crucial.
However, considering that magnetic fields have detrimental effects on both topological insulators and superconductors, we are motivated to keep time-reversal invariance in the system and to avoid magnetic fields altogether. This requires a different approach where the competition between Zeeman gap and superconducting proximity gap is replaced by a different mechanism, as we outline next.

The proximity to an $s$-wave superconductor induces superconducting pairings between edge modes that have opposite momenta and spins. Importantly for us, there are two competing superconducting mechanisms. The first one is the usual  pairing  that corresponds to a process where a Cooper pair tunnels as a whole in one of topological insulators.
This process induces a coupling between right and left movers that also have opposite spins as a consequence of time-reversal invariance.\cite{Hasan_review} The second mechanism is based on a non-local or `crossed Andreev'  pairing\cite{Feinberg_2000,Recher_Sukhorukov_Loss,Patric_TI_2,Recher_Loss,Bena,Schonenberger,Heilblum,Andreev_Yaroslav,Daniel_Yaroslav_PRL_TI_CAS}
and corresponds to the case when a Cooper pair gets split between two topological insulators: One electron from the Cooper pair tunnels into one of the topological insulators, while its partner tunnels into the other one. 

The crossed Andreev  pairing dominates over the usual one in the regime of strong electron-electron interactions if the separation between edge states of neighboring topological insultors  does not exceed the coherence length and the Fermi wavelength of the superconductor.\cite{Recher_Loss} If the edges are separated by a large distance, the crossed Andreev pairing gets suppressed, and the usual one dominates. All this allows us to create interfaces between two different types of induced superconductivity where the two gap mechanisms compete. In such a case, one can expect the emergence of zero-energy bound states.

Indeed, we demonstrate that  the crossed Andreev and usual  pairing also compete with each other in the opening of gaps in the energy spectrum. Importantly, if all pairings are of the same strength, the system is gapless. This signals the existence of two topologically distinct phases. As a consequence, there are zero-energy bound states at the interfaces between regions dominated either by the crossed Andreev  or by the usual pairing. Due to the preserved time-reversal symmetry, these bounds states are Kramers pairs of Majorana fermions for standard topological insulators, while they are pairs of parafermions for fractional topological insulators which carry edge states that obey fractional statistics.\cite{Ady_FTI}

We consider two setups that both are based on the competition between local and non-local pairing mechanisms. However, they differ by the direction of edge mode propagation in each topological insulator (or equivalently by the directions of spin orbit vectors). In our first, {\it para-helical}, setup  considered in Sec. II, edge modes with the same spins propagate in the same direction along their common boundary in both topological insulators. In contrast to that,  in our second, {\it ortho-helical}, setup   considered in Sec. III, edge modes with opposite spins  propagate in the same direction along their common boundary. In spite of this difference, our main result on the presence of zero-energy bound states with non-Abelian statistics holds for both setups. 

\section{Para-helical setup}

\begin{figure}[t!]
\includegraphics[width=\linewidth]{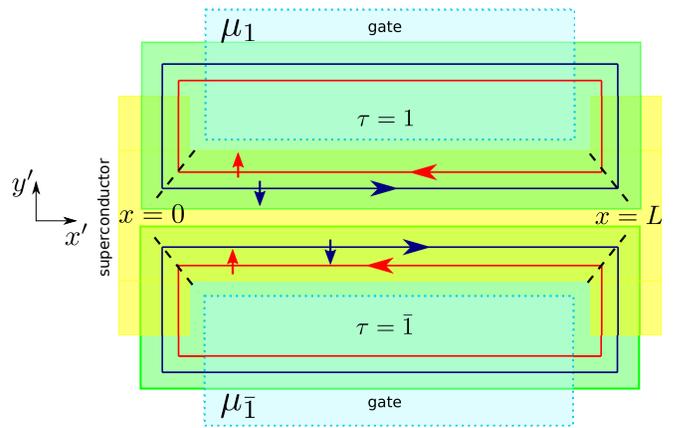}
\caption{Para-helical setup. Sketch of two topological insulators (green rectangles, labeled by $\tau=1,{\bar 1}$) that are tunnel-coupled to an underlying $s$-wave superconductor (yellow) which induces superconducting proximity effects in the TIs. The chemical potentials, controlled by the gates (light blue), are {\it the same} in the two samples, $\mu_1=\mu_{\bar 1}$. The helical edge modes (blue and red lines with arrows) in the TIs circulate in  {\it opposite}  directions due to  opposite spin orbit interaction vectors. The crossed Andreev  pairing dominates over the usual  pairing along the common boundary of the two TIs (along the $x'$ direction). In contrast to that, 
along the $y'$ direction the dominant pairing  
is the usual one produced by Cooper pairs tunneling into each of the TIs as a whole. The propagation direction of spin-downs defines the $x$ axis.
The interfaces  between two such regions at $x=0$ and at $x=L$  (shown by dashed lines) host  the zero-energy bound states such as Kramers pairs of Majorana fermions or Kramers pairs of parafermions  depending on the nature of the TI (see text).}
\label{fig:SOI}
\end{figure}

\subsection{Model Hamiltonian}

\begin{figure}[tb!]
\includegraphics[width=0.8\linewidth]{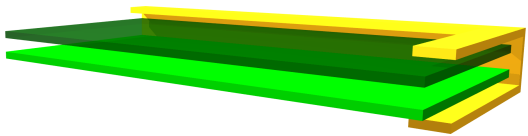}
\caption{Alternative realization of the para-helical setup. A bilayer consists of two identical topological insulators (dark and light green slabs) tuned at the same chemical potential. The shape of an $s$-wave superconductor (yellow L-shaped slab) is designed such that the crossed Andreev pairing (coupling states from different TIs) dominates along one of the edge (here, along the long edge), whereas the usual superconducting pairing (coupling states from the same TI) dominates at the other edge (here, along the short edge). Again, there are two interfaces separating the two regions, each of them
containing Kramers pairs of bound states. Note that while at one interface (located at the corner of the L-shaped slab) the bound states are well-localized, there are spread out over the normal region edge states forming the other interface.  }
\label{fig:TI_bilayer}
\end{figure}

 The first setup we consider consists of two two-dimensional topological insulators (TIs),\cite{Hasan_review,Volkov_TI1,Volkov_TI2,Volkov_TI3,Fu_Kane,Zhang_TI} labeled by $\tau=1$ and  $\tau={\bar 1}(\equiv -1)$,
that are brought into proximity to an $s$-wave superconductor, see Fig. \ref{fig:SOI}. The chemical potentials $\mu_\tau$, are assumed to be equal, $\mu_1=\mu_{\bar 1}$,  and tuned inside the band gaps
 of the topological insulators.
 The topologically protected edge modes
propagate along the edges of the system in the direction determined by the spin orbit interaction (SOI) vector. In this section, we consider systems in which modes with the same spin direction propagate in the same direction along their common boundary. This can be achieved by placing two identical TIs on top of each other separated by a thin insulating layer, see Fig. \ref{fig:TI_bilayer}. Alternatively, we can consider schemes involving samples with opposite signs of spin orbit interaction in the two TIs [for example, 1) by growing two samples with reversed order of layers in the heterostructure, 2) by flipping one of two identically grown samples], see Fig.~\ref{fig:SOI}.
Thus,  edge modes with spin up propagate, say, in the clockwise (anticlockwise) direction in the upper (lower) system  $\tau=1$ ($\tau=\bar 1$), where we choose the spin quantization axis to be along the $z$ axis  perpendicular to the plane of  the 2D system, see Fig.~\ref{fig:SOI}.
As a consequence, both spin-up (down) modes propagate to the right (left) along their common boundaries. Further, we refer to this setup as the 
{\it para-helical} one (since the spin orbit vectors are opposite to each other).

The kinetic part of the Hamiltonian is written as
\begin{align}
H_0 = - i \hbar \upsilon_F \sum_{\tau} \int dx  [R^\dagger_\tau \partial_x R_\tau- L^\dagger_\tau \partial_x L_\tau ],
\label{ho}
\end{align}
where $ \upsilon_F$ is the Fermi velocity. The slowly-varying right (left) mover field of the $\tau$ system $R_\tau (x)$ [$L_\tau (x)$] are defined around the Fermi points $\pm k_F$ that are determined by the position of the chemical potential $\mu_\tau$. Here, we choose the $x$ axis to be always aligned with the  propagation direction of the spin-down edge modes.

Next, we explore all possible proximity induced superconducting pairings taking into account that they should couple modes with opposite momenta and opposite spins.

\begin{figure}[b!]
\includegraphics[width=0.5\linewidth]{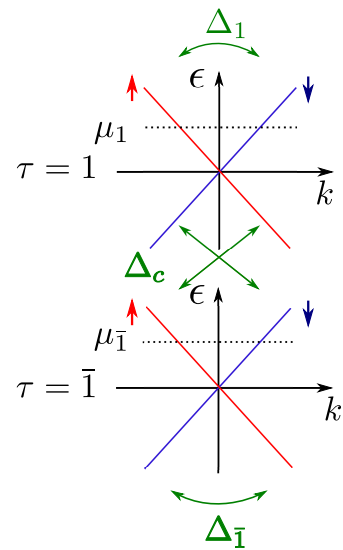}
\caption{The spectrum of two pairs of edge modes in topological insulators labeled by $\tau$ corresponding to the para-helical setups shown in  Figs.~\ref{fig:SOI} and \ref{fig:TI_bilayer}. 
The proximity induced pairing amplitude $\Delta_\tau$ induces coupling between edge modes from the same TI, resulting in proximity gaps at the Fermi level in each TI separately.
If the chemical potentials $\mu_\tau$ are the same in the two samples, $\mu_{1}=\mu_{\bar 1}$,
 the crossed Andreev pairing with amplitude $\Delta_c$  couples a spin-up state at the Fermi level in the upper TI ($\tau=1$, red arrow) with a spin-down state at the Fermi level in the lower TI 
 ($\tau={\bar 1}$, blue arrow), and vice versa.
 The crossed pairing is dominant in the regime of strong electron-electron interactions since it costs more energy to insert two charges simultaneously into the same edge of a given TI than each of them separately into different TIs.
}
\label{spectrum_SOI}
\end{figure}

 First,  Cooper pairs that tunnel together as a whole  into each of the two systems lead to the usual proximity induced superconductivity of  strength $\Delta_\tau$ dependent on the tunnel coupling between the superconductor and the TIs, see Fig. \ref{spectrum_SOI}.  The corresponding pairing Hamiltonian describing this usual proximity pairing takes  the form
\begin{align}
H_{\tau} =  \int dx\ &\Big[ \frac{\Delta_{\tau}}{2} (R_\tau^\dagger L^\dagger_{\tau} - L^\dagger_{\tau} R_{\tau}^\dagger) 
+H.c.\Big].
\label{htau}
\end{align}

\begin{figure}[tb!]
\includegraphics[width=0.8\linewidth]{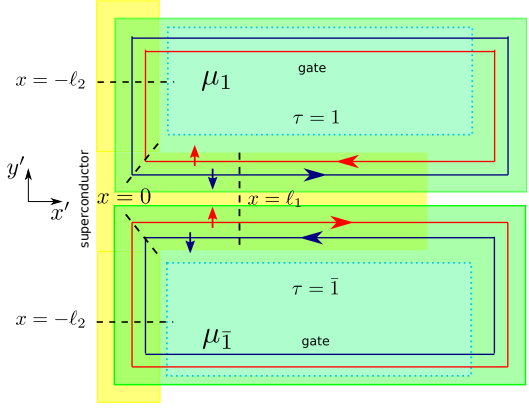}
\caption{Alternative realization of the para-helical setup based on a spatial decay of the crossed Andreev pairing. The setup consists of two identical topological insulators (green slabs) tuned at the same chemical potential. The shape of the $s$-wave superconductor (yellow T-shaped slab) is designed such that the crossed Andreev pairing (coupling states from different TIs) dominates around $x=0$, where the distance between the samples is smallest, and decays away from this point, giving  way to the usual superconducting pairing (coupling edge states from the same TI). The decay along the $y'$ axis is determined by the coherence length and the Fermi wavelength of the superconductor. In contrast, the decay along the $x'$ axis is much faster and arises due to a momentum mismatch between the spin-up helical edge states of one TI and the spin-down helical edge states of the other TI. As a result, there are two interfaces between different superconducting pairing regimes at $x=-\ell_2$ and $x=\ell_1$.}
\label{fig:TI_alternative}
\end{figure}

Second, in addition to the usual proximity effect induced in each of the two systems separately, there is a crossed (or split) Andreev process that induces a  superconductivity of  strength $\Delta_c$ in both systems simultaneously. Such superconducting pairing corresponds to a process when Cooper pairs get split so that one partner of the pair tunnels into one system and its partner into the other one. Consequently, such a split pairing dominates in the regime of strong electron-electron interactions since it costs less energy to insert only one electron into one edge state compared to inserting two electrons into the same edge state. \cite{Recher_Sukhorukov_Loss,Recher_Loss,Bena} One should also note that the crossed Andreev  pairing strength $\Delta_c$ gets exponentially suppressed if the distance between the edge states of the upper and the lower TIs $d$ is exceeds  the  coherence length of the superconductor.\cite{Recher_Loss,Sasha,Daniel_Yaroslav_PRL_TI_CAS}  Moreover, $\Delta_c$ is suppressed as an 
inverse power law of $(k_{Fsc}d)$, where $k_{Fsc}$ is the Fermi wavevector of a two- or three-dimensional bulk superconductor. This suppression can be overcome, however, by using an intermediate layer of proximity induced superconductor with smaller $k_{Fsc}$. \cite{Schonenberger} 
In addition, such a spatial dependence of the crossed Andreev process enables us to envisage a third variation of the para-helical setup, see Fig. \ref{fig:TI_alternative}. This setup consists of two identical topological insulators with equal chemical potentials brought into proximity to a T-shaped superconductor. The crossed Andreev term is the strongest around $x=0$ away from which it decays and gives  way to the usual superconducting pairing at $x<-\ell_2$ and $x>\ell_1$. As a result, we again arrive at interfaces between two regions dominated by different pairing mechanisms. In what follows we focus on the setup depicted in Fig. \ref{fig:SOI} due to the simplicity of treatment of uniform pairings. However, we note that the main results, such as the presence of bound states, also hold for the non-uniform case.

The corresponding Hamiltonian for the proximity effect due to crossed Andreev superconducting pairing takes the form
\begin{align}
H_{c} =\int dx\ &\Big[ \frac{\Delta_{c}}{2} (R_1^\dagger L^\dagger_{\bar 1}+R_{\bar1}^\dagger L^\dagger_{1} - L^\dagger_{\bar 1} R_{ 1}^\dagger-L^\dagger_{1} R_{\bar 1}^\dagger) 
+H.c.\Big],
\label{hc}
\end{align}
and describes the pairing coupling between right movers with spin up from one system with left movers with spin down from the other system, see Fig. \ref{spectrum_SOI}.

Both crossed Andreev and usual superconducting pairings open a full gap in the spectrum. As a result, one can expect a competition between two pairing mechanisms.  From now on, we focus on the interfaces between two regions, see Fig. \ref{fig:SOI}. In one region, the dominating superconducting pairing arises from the crossed Andreev term $H_c$; in the other region, the crossed Andreev superconducting pairing is suppressed, so the edge modes are gapped only by the usual superconducting terms $H_{\tau}$. Such interfaces can potentially host exotic bound states (with non-trivial braiding statistics), which is the focus of the following sections.

\subsection{Kramers Pairs of Majorana Fermions}

In this subsection, we consider an effective non-interacting single particle model of topological insulator edge modes with excitations possessing an elementary electron charge $e$.\cite{Hasan_review,Volkov_TI1,Volkov_TI2,Volkov_TI3,Fu_Kane,Zhang_TI} This allows us to work within the fermionic Nambu representation in the basis of eight fermion operators,\cite{Klinovaja2012,Klinovaja_Loss_Ladder,Rotating_field,Braunecker}
$\Psi=(R_1, L_1, R_{\bar 1}, L_{\bar 1}, R_1^\dagger, L_1^\dagger, R_{\bar 1}^\dagger, L_{\bar 1}^\dagger)$. The Hamiltonian density $\mathcal{H}$, defined via $H=(1/2)\int dx\ \Psi(x)^\dagger \mathcal{H} \Psi(x)$ and the corresponding total Hamiltonian $H=H_0+H_1+H_{\bar 1}+H_c$  [see Eqs.~(\ref{ho})-(\ref{hc})], becomes written in terms of Pauli matrices, 
\begin{align}
&\mathcal{H}= \hbar \upsilon_F \hat k \rho_3  - \frac{\Delta_{1}}{2} (1+\tau_3) \rho_2 \eta_2 \nonumber \\
&\hspace{83pt}- \frac{\Delta_{\bar 1}}{2} (1-\tau_3) \rho_2 \eta_2 
 - \Delta_c \eta_2 \tau_1 \rho_2,
 \label{hamiltoniandensity1}
\end{align}
where $\hbar \hat k$ is the momentum operator.
Here, the Pauli matrices $\rho_i$ ($\eta_i$) act in  left-right mover (electron-hole) space, and the Pauli matrices $\tau_i$ act in  sample space. In addition, without loss of generality, we assume that $\Delta_\tau$ and $\Delta_c$ are  non-negative. The energy spectrum of  Eq.~(\ref{hamiltoniandensity1}) is given by
\begin{align}
&E_{\pm,\pm}= \pm {\Big [} \Big(\Delta_{1}^2 + \Delta_{\bar 1}^2 +2  \Delta_{c}^2 + 2 (\hbar \upsilon_F k)^2   \\
&\hspace{40pt}\pm (\Delta_{1} + \Delta_{\bar 1}) \sqrt{(\Delta_{1} - \Delta_{\bar 1})^2 + 4\Delta_c^2}\Big)/2 {\Big ]}^{1/2}\nonumber,
\end{align}
where $k$ is the momentum and each level $E_{\pm,\pm}$ is twofold degenerate (at fixed $k$), which is a direct consequence of the assumed time-reversal invariance of the system. 
The system gap 
is defined by the energy difference at $k=0$,
{\it i.e.},  $2|\Delta_{k=0}|\equiv E_{+,-}(k=0) -E_{-,-}(k=0)$, 
which gives explicitly
\begin{align}
\Delta_{k=0}=\Big[\Delta_{1} + \Delta_{\bar 1}-\sqrt{(\Delta_{1} - \Delta_{\bar 1})^2 + 4\Delta_c^2}\Big]/2. 
\label{gap1}
\end{align}
From this we see immediately  that the spectrum is gapless if $\Delta_c^2=\Delta_1\Delta_{\bar1}$, while it is   fully gapped otherwise. The closing and reopening of the gap as a function of system parameters could signal a topological phase transition that separates two superconducting phases: the topological one with $\Delta_{k=0}<0$ and the trivial one with $\Delta_{k=0}>0$. If an interface between such two phases (two regions in real space) is created, it should host localized zero energy states. 
In the first region the gap is dominated by, say, the crossed Andreev superconductivity $\Delta_c$ with  $\Delta_{k=0}<0$. In the second region, the gap is determined by the usual pairing $\Delta_1$ and $\Delta_{\bar 1}$, respectively, with $\Delta_{k=0}>0$.

As an example, we consider such an interface at $x=0$ separating two regions characterized by different strengths of the crossed Andreev pairing amplitude: $\Delta_c >\sqrt{\Delta_1\Delta_{\bar1}}>0$ for $x>0$ and $\Delta_c=0$ for $x<0$. The usual  pairing amplitudes $\Delta_\tau$ are assumed to be uniform for the sake of simplicity. Indeed, at this interface, we find two zero-energy bound states $\Psi_{MF1}$ and $\Psi_{MF\bar1}$ localized at $x=0$. These two states are Majorana fermions, $\hat\Psi_{MFi}=\hat\Psi_{MFi}^\dagger$, and they are Kramers partners, $\Psi_{MF1}=T\Psi_{MF\bar1}$. Here, $T=i\sigma_2 K$ is the time reversal operator composed of $K$, the complex conjugation operator [$\Psi^*(x)=K\Psi(x)$], and of the Pauli matrix $\sigma_i$ acting on spin.  In general, a Kramers pair of Majorana fermion wavefunctions can be written in the basis $\tilde \Psi=(\Psi_{1,\downarrow}, \Psi_{1,\uparrow}, \Psi_{\bar 1,\downarrow}, \Psi_{\bar 1,\uparrow},\Psi_{1,\downarrow}^\dagger, \Psi_{1,\uparrow}^\dagger, \Psi_{\bar 1,\downarrow}^\dagger, \Psi_{\bar 1,\uparrow}^\dagger)$ as
\begin{align}
\Psi_{MF1}(x)=\begin{pmatrix}
f_1(x)\\g_1(x)\\f_{\bar 1}(x)\\g_{\bar 1}(x)\\f_1^*(x)\\g_1^*(x)\\f_{\bar 1}^*(x)\\g_{\bar 1}^*(x)
\end{pmatrix},\ \ 
\Psi_{MF\bar1}(x)=\begin{pmatrix}
g^*_1(x)\\-f^*_1(x)\\g^*_{\bar 1}(x)\\-f_{\bar 1}^*(x)\\g_1(x)\\-f_1(x)\\g_{\bar 1}(x)\\-f_{\bar 1}(x)
\end{pmatrix},
\end{align}
where the complex functions $f_{\tau }(x)$ and $g_{\tau}(x)$ are, in principle, arbitrary up to the normalization prefactor which will be suppressed in the following. The annihilation (creation) operators $\Psi_{\tau,\sigma}$ ($\Psi_{\tau,\sigma}^\dagger$) act on electrons with spin $\sigma=\uparrow, \downarrow$ located in the $\tau$ wire. We note that wavefunctions in the basis $\tilde \Psi$ can be easily obtained from wavefunctions written in the basis $\Psi$  by restoring the fast oscillating factors $e^{\pm ik_Fx}$.

In the case of a sharp drop of the crossed Andreev  pairing amplitude $\Delta_c$ at $x=0$ specified above, these functions are obtained straightforwardly~\cite{Klinovaja2012} and given by
\begin{align}
&f_1 = - i g_1^*= s_1\frac{\Delta_1-\Delta_{\bar 1}-\sqrt{(\Delta_1-\Delta_{\bar 1})^2+4\Delta_c^2}}{2\Delta_c} ,\nonumber\\
&f_{\bar 1} = - i g_{\bar 1}^* = s_{\bar 1},\nonumber\\
      &s_\tau=e^{-ik_F x} \times\begin{cases} 
      e^{-x/\xi_c},& x>0\\
     - e^{x/\xi_{\tau}}, & x<0\, .
        \end{cases} \,\, 
\end{align}
The localization lengths $\xi_c$ and $\xi_{\tau}$ are inverse proportional to the gaps opened at the Fermi points and are explicitly given by
\begin{align}
&\xi_c=2\hbar \upsilon_F/(\sqrt{(\Delta_1-\Delta_{\bar 1})^2+4\Delta_c^2}-\Delta_1-\Delta_{\bar 1}),\\
&\xi_{\tau} = \hbar\upsilon_F/\Delta_{\tau}.
\end{align}

Above we have considered only the interface at $x=0$. An analogous interface also occurs at the other end,  $x=L$, of the  region dominated by crossed Andreev pairing  (see Fig. \ref{fig:SOI}). By analogy, we can show that it hosts also a Kramers pair of Majorana fermions. If the length of the crossed Andreev dominated region is much longer than the localization length, $L\gg \xi\equiv {\rm max} \{1/\xi_c, 1/\xi_\tau \}$, the wavefunctions of the MFs localized at two different interfaces do not overlap, so the MFs can be considered as independent. If $L$ gets comparable with the localization length $\xi$, two overlapping MFs hybridize into a complex fermionic state whose energy is generally non-zero.\cite{diego} We note that the two MFs of the same
Kramers pair are located at the same position but being orthogonal states do not hybridize as long as time-reversal symmetry is present.\cite{TRI1,TRI2,TRI3,TRI4,TRI5,TRI6,TRI7,TRI8,TRI9,TRI10,PFs_Loss_2} If this symmetry is broken, e.g. by an external magnetic field or by magnetic impurities, then they can also split in energy away from zero and become a complex fermionic state.

\subsection{Kramers Pairs of Parafermions in Fractional Topological Insulators}

So far we have worked with an effective single particle Hamiltonian, which is appropriate in the regime considered. In this subsection, we focus on the regime where electron-electron interactions dominate and work with fractional topological insulators in which elementary excitations possess an effective charge $e/m$ with $m$ being a positive odd integer.\cite{Ady_FTI,Footnote} 

To analyze this problem we make use of the bosonization formalism.
In terms of chiral bosonic fields $\phi_{r\tau}$, the electron operators are then rewritten as $R_\tau=e^{im\phi_{1\tau}}$ and $L_\tau=e^{im\phi_{\bar1\tau}}$.  By analogy, we introduce chiral bosonic fields $\tilde\phi_{r\tau}$ describing hole operators in Nambu space. To satisfy the anticommutation relations between original fermionic operators, we work with the following non-zero commutators for the bosonic fields,
\begin{align}
&[\phi_{r\tau}(x),\phi_{r'\tau'}(x')]=\frac{i\pi r}{m} \delta_{rr'}\delta_{\tau \tau'} {\rm sgn} (x-x'),\label{com1}\\
&[\tilde\phi_{r\tau}(x),\tilde\phi_{r'\tau'}(x')]=\frac{i\pi r}{m} \delta_{rr'}\delta_{\tau \tau'} {\rm sgn} (x-x').\label{com2}
\end{align}
All other commutators are assumed to vanish.

To proceed, we bosonize the Hamiltonians 
$H_\tau$  [see Eq.~(\ref{htau})] and $H_c$  [see Eq.~(\ref{hc})],
\begin{align}
&H_c =\Delta_c \big(\cos [m(\phi_{11}-\tilde{\phi}_{\bar 1 \bar1})]+  \cos [m(\phi_{1\bar1}-\tilde\phi_{\bar 1 1})]\nonumber\\
&\hspace{30pt}-\cos [m(\phi_{\bar 1 \bar1}-\tilde{\phi}_{ 1 1})]-\cos [m(\phi_{\bar 1 1}-\tilde{\phi}_{ 1 \bar 1})]\big),\\
&H_1 =\Delta_1 \big(\cos [m(\phi_{11}-\tilde\phi_{\bar 1 1})]-\cos [m(\phi_{\bar 11}-\tilde\phi_{1 1})]\big),\\
&H_{\bar 1} =\Delta_{\bar 1}\big(\cos [m(\phi_{1\bar1}-\tilde\phi_{\bar 1 \bar1})]-\cos [m(\phi_{\bar1\bar1}-\tilde\phi_{ 1 \bar1})]\big).
\end{align}
Furthermore, in  Nambu representation one can easily see that the total Hamiltonian can be separated into two independent (commuting) parts, $H^{(1)}$ and $H^{(\bar 1)}$, so that $H=H^{(1)}+H^{(\bar 1)}$. Here, $H^{(1)}$ [$H^{(\bar 1)}$] act in the subspace composed of $\phi_{1\tau}$ and $\tilde{\phi}_{\bar 1 \tau}$ ($\phi_{\bar 1\tau}$ and $\tilde{\phi}_{ 1 \tau}$). These two parts are connected by time-reversal symmetry, so their eigenstates are Kramers partners of each other. The non-quadratic part of $H^{(1)}$ consists of $H_1^{(1)}$, $H_{\bar 1}^{(1)}$, and  $H_c^{(1)}$ given by
\begin{align}
&H_c^{(1)} =\Delta_c \big(\cos [m(\phi_{11}-\tilde{\phi}_{\bar 1 \bar1})]\nonumber\\
&\hspace{100pt}+  \cos [m(\phi_{1\bar1}-\tilde\phi_{\bar 1 1})]\big),\label{hc1}\\
&H_1^{(1)} =\Delta_1 \cos [m(\phi_{11}-\tilde\phi_{\bar 1 1})], \label{h11}\\
&H_{\bar 1}^{(1)} =\Delta_{\bar 1}\cos [m(\phi_{1\bar1}-\tilde\phi_{\bar 1 \bar1})] \label{h12}.
\end{align}
From now on, we focus on $H^{(1)}$, keeping in mind that the eigenstates of $H^{(\bar 1)}$ can be easily obtained by acting with the time-reversal  operator $T$ on the eigenstates of $H^{(1)}$.

In a next step, we express the chiral fields $\phi_{1\tau}$ and $\tilde{\phi}_{\bar 1 \tau}$ in terms of the conjugated $\phi_\eta$ and $\theta_\eta$ fields, $\eta=\pm 1$, defined as
\begin{align}
&\phi_1 = m (\phi_{11}+\phi_{1\bar1}-\tilde\phi_{\bar11}-\tilde\phi_{\bar1\bar1})/2,\\
&\theta_1 = m (\phi_{11}+\phi_{1\bar1}+\tilde\phi_{\bar11}+\tilde\phi_{\bar1\bar1})/2,\\
&\phi_{\bar1} =  (\phi_{11}-\phi_{1\bar1}-\tilde\phi_{\bar11}+\tilde\phi_{\bar1\bar1})/2,\\
&\theta_{\bar 1} = (\phi_{11}-\phi_{1\bar1}+\tilde\phi_{\bar11}-\tilde\phi_{\bar1\bar1})/2,
\end{align}
or inverting
\begin{align}
&\phi_{1\tau}=\left(\frac{\phi_1+ \theta_1 }{m} + \tau \phi_{\bar 1} +\tau \theta_{\bar 1}\right)/2,\\
&\tilde\phi_{\bar1\tau}=\left(\frac{-\phi_1+ \theta_1 }{m} - \tau \phi_{\bar 1} +\tau \theta_{\bar 1}\right)/2.
\end{align}

As a result, $H_c^{(1)}$ and $H_\tau^{(1)}$ [see Eqs. (\ref{hc1})-(\ref{h12})] become in the new basis 
\begin{align}
&H_c^{(1)} =2 \Delta_c \cos (\phi_1) \cos(m\theta_{\bar 1}),\label{hcb}\\
&H_1^{(1)} =\Delta_1 \cos (\phi_1 + m\phi_{\bar 1}),\\
&H_{\bar 1}^{(1)} =\Delta_{\bar 1} \cos (\phi_1 - m\phi_{\bar 1}).\label{h2b}
\end{align}

If there was only one interaction term present in the Hamiltonian, for example $H_c^{(1)}$ ($H_\tau^{(1)}$), and if it was relevant (in a renormalization group sense), then the corresponding coupling constant $\Delta_c$ ($\Delta_\tau$) would grow (under renormalization flow due to interactions) leading to a gapped spectrum. In addition, the combination of fields standing under the corresponding cosine would get pinned to a constant such that the total energy is minimized. \cite{Giamarchi}

However, if both terms $H_c^{(1)}$  and $H_\tau^{(1)}$ are present, only one of them can be ordered (so that the corresponding fields get pinned), as the crossed Andreev term $H_c$ and the usual pairing term $H_\tau$ do not commute with each other. This is easily seen from Eqs. (\ref{htau}) and (\ref{hc})  written in the fermionic representation. However, one can arrive at  the same conclusion also in the bosonic representation [see Eqs. (\ref{hcb})-(\ref{h2b})] by using the commutation relations for the new fields $\phi_\eta$ and $\theta_\eta$ obtained from Eqs. (\ref{com1}) and (\ref{com2}):
\begin{align}
&[\phi_{\bar 1} (x),\theta_{\bar 1} (x')]=\frac{i\pi }{m} {\rm sgn} (x-x'), \nonumber \\
&[\phi_1 (x),\theta_{\bar 1} (x')]=[\phi_1 (x),\phi_{\bar 1} (x')]=0,
\label{conjugate_commutations1}
\end{align}
where, in addition, each of the bosonic fields commutes with itself.
As a consequence, the two different  pairing terms do not commute and thus cannot be ordered simultaneously within the same space region. To avoid this situation, in what follows we thus work in the regime of strong electron-electron interactions such that $H_c$ dominates over $H_{\tau}$ if both of them are present, and pinning is possible. Interaction effects favor energetically the separation and injection of the two electrons
from a Cooper pair into separate edges over a simultaneous injection of the two electrons as a whole into the same edge.

Similarly, to the effective single-particle case considered in the previous subsection, we focus on the interface separating the region dominated by crossed Andreev pairing from the regions 
where the usual pairing dominates, see Fig. \ref{fig:SOI}.
The crossed Andreev term $H_c$ is dominant inside the interval $0<x<L$ which is surrounded by two regions,  $x<0$ and $x>L$, where the usual superconducting terms $H_\tau$ are dominant.

\begin{figure}[b!]
\includegraphics[width=\linewidth]{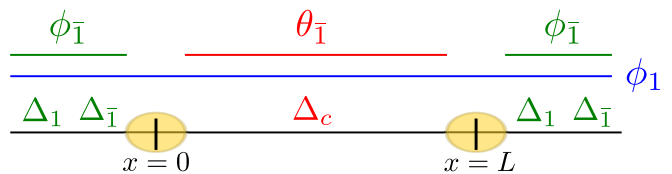}
\caption{Pinning of bosonic fields in the regime of strong electron-electron interactions. The field $\phi_1$ (blue line) is pinned uniformly in the entire system. The field $\phi_{\bar 1}$ ($\theta_{\bar 1}$) is pinned in the region where the usual  pairing term $\Delta_\tau$ (crossed Andreev  pairing term $\Delta_c$) is dominant, indicated by green (red) lines. 
The different regions are separated by interfaces of size $\ell$ located at $x=0$ and $x=L$ (yellow circles). 
Within these interfaces  the non-commuting fields $\phi_{\bar 1}$ and  $\theta_{\bar 1}$
overlap and fluctuate, giving rise to zero-energy bound states that can be identified as Kramers pairs of parafermions (one pair at each interface).
}
\label{pinning}
\end{figure}

Deep inside the region $0<x<L$ [$x<0$ or $x>L$], the corresponding bosonic fields $\phi_1$ and $\theta_{\bar 1}$ ($\phi_1$ and $\phi_{\bar 1}$) are pinned as to minimize the total energy, see Fig. \ref{pinning}.  First, we note that the bosonic field $\phi_1$ should be pinned uniformly in the entire system, $\phi_1 =\pi \hat M$. Second, the other fields are pinned as
\begin{align}
&\phi_{\bar 1} =\begin{cases}
                 \pi ( \hat M + 1+2 \hat l_{\bar 1})/m, & x<0 ,\\
                  \pi ( \hat M + 1+2 \hat l_{1})/m, & x>L,
                \end{cases} \\
                &\theta_{\bar 1}=\pi(\hat M+1+2\hat n)/m, \ x\in(0,L).
\end{align}
Here, $\hat M$, $\hat n$, and $\hat l_{\tau}$ are integer-valued operators. The commutation relations between them follow from Eqs.~(\ref{conjugate_commutations1}) and are given by
\begin{align}
&[\hat l_1,\hat n]=-[\hat l_{\bar 1},\hat n]=-\frac{i m}{4\pi },\label{lcom}\\
&[\hat M,\hat n]=[\hat M,\hat l_1]=[\hat M,\hat l_{\bar 1}]=0 \label{Mcom}.
\end{align}
As a result of the two fields being pinned simultaneously ($\phi_1$ and $\theta_{\bar 1}$ in the crossed Andreev region; $\phi_1$ and $\phi_{\bar 1}$ in the usual pairing region), the energy spectrum away from interfaces is fully gapped. In other words, the system does not host zero-energy states in the bulk. \cite{PF_Mong,PF_Clarke}

Importantly, while two non-commuting fields cannot be pinned simultaneously, at interfaces between two regions both $\phi_{\bar 1}$ and $\theta_{\bar 1}$ are fluctuating inside the interface region of size $\ell$ and the bulk gap can close (apart from finite size level spacings). \cite{PF_Clarke}
Consequently, the interfaces might host zero-energy states with the localization length determined by $\ell$. To identify these states, we define the operator $\alpha_1$ ($\alpha_{\bar 1}$) at the interface $x=0$ ($x=L$):
\begin{align}
&\alpha_1 = e^{i \frac{2\pi}{m}  (\hat n+\hat l_1) -i\frac{\pi}{2}},\ \  \alpha_{\bar 1} = e^{i \frac{2\pi}{m}  (\hat n+\hat l_{\bar 1}) +i\frac{\pi}{2}},
\end{align}
These operators are easily seen to commute with the total Hamiltonian, and thus they represent zero-energy bound states.\cite{PF_Clarke}

Next, we identify the properties of $\alpha$ operators using derived  above commutators between the integer-valued operators, see Eqs. (\ref{lcom}) and (\ref{Mcom}). First, we note that $\alpha_{\tau}^m=1$. Second, we  obtain for the commutation relations between two different operators
\begin{align}
&\alpha_1 \alpha_{\bar 1}  = \alpha_{\bar 1} \alpha_1 e^{2\pi i/m}.
\end{align}
These operators thus possess the same properties as ${\mathbb Z}_m$ parafermion operators.\cite{}

We also identify the degeneracy of the ground state. The operator $\alpha_1^\dagger \alpha_{\bar 1} = -e^{-i\pi/m}e^{2\pi i (\hat l_{\bar 1} - \hat l_1)/m}$ has the property that $(\alpha_1^\dagger \alpha_{\bar 1})^m=1$. Its $m$ eigenstates denoted as $\ket q$ correspond to the eigenvalues $e^{2\pi i q/m}$. Thus, the ground state of $H^{(1)}$ and  $H^{(\bar 1)}$ is $m$-fold degenerate.\cite{PF_Clarke,PF_Mong} Taking into account Kramers degeneracy, we find that the ground state of the system described by $H=H^{(1)}+H^{(\bar 1)}$ is represented by $m$ Kramers pairs of parafermions.

\section{Ortho-helical setup}

\begin{figure}[!tb]
\includegraphics[width=\linewidth]{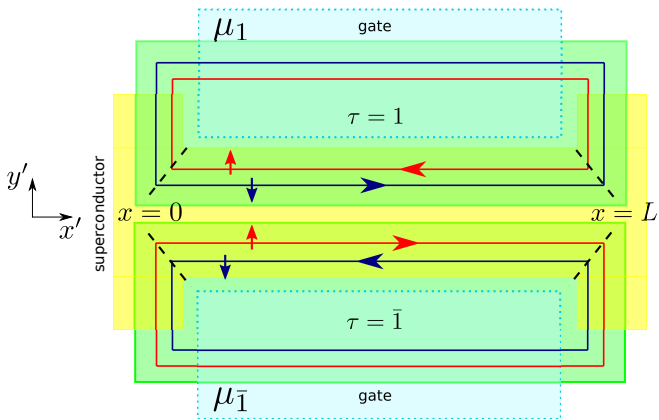}
\caption{Ortho-helical setup. Sketch of two topological insulators ($\tau=\pm 1$, green rectangles) brought into proximity to a superconductor (yellow). The chemical potentials, controlled by the gates (light blue), are tuned to be {\it opposite} in the two samples, $\mu_1 = -\mu_{\bar 1}$.  The edge modes circulate in {\it the same} direction in both samples as their spin orbit interaction vectors are codirectional.
As a result of strong electron-electron interactions, the crossed Andreev superconducting pairing dominates over the usual superconducting pairing along the common boundary of the two samples ($x'$ direction).
In contrast to that, in the region where the edge states propagate along the $y'$ direction, the only present superconducting pairing is the usual one produced by Cooper pairs tunneling into one of the samples as a whole.}
\label{same_soi}
\end{figure}

\subsection{Model Hamiltonian}

The model considered in this section is similar to the one considered above in Sec. II. The main ingredients are the same: the two topological insulators,\cite{Hasan_review,Volkov_TI1,Volkov_TI2,Volkov_TI3,Fu_Kane,Zhang_TI,Ady_FTI} again labeled by $\tau=1,{\bar 1}$, are brought into the proximity to an ordinary $s$-wave superconductor. However, in contrast to the previous case,
the spin orbit vectors of the two TIs are both aligned in the positive direction along  the $z$ axis. This means that in both systems spin up (down) states propagate in the clockwise (anticlockwise) direction, see Fig. \ref{same_soi}. We refer to this case as to the {\it ortho-helical} setup (since the spin orbit interaction vectors point along the same direction).

 Alternatively, one can also aim at using pairs of edge modes localized at the opposite edges of a {\it single} topological insulator. However, such a setup poses some additional difficulties for implementations. First,
it is challenging to keep the chemical potentials at opposite values $\mu_1=-\mu_{\bar 1}$ at  opposite edges in the {\it same} sample where all edge states are connected. So, the setup should be tuned close to the Dirac point $\mu_1\approx-\mu_{\bar 1}\approx 0$, which, however, requires very precise control. Second, a compromise for the sample width needs to be found. On one hand, the edge modes from opposite edges separated by $d$ should not overlap, thus  $d$  should be much larger than the penetration length of edge states inside the sample. On the other hand, the separation $d$ should be smaller than the coherence length  of the superconductor in order for the crossed Andreev  pairing to be dominant. All this makes  an experimental realization using a single TI presumably more challenging  than using two of them.

We emphasize that for notational simplicity we use the same symbols for fermionic and bosonic operators and Hamiltonians as in the previous section. We believe that this should not lead to ambiguities, as long as we clearly distinguish between ortho-helical and para-helical setups.

\begin{figure}[!t]
\includegraphics[width=0.5\linewidth]{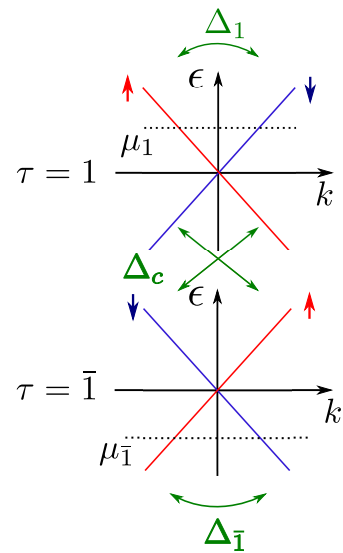}
\caption{The spectrum of two  pairs of edge modes in topological insulators  corresponding to the ortho-helical setup shown in Fig. \ref{same_soi}. The superconducting pairing is similar to the one shown in Fig. \ref{spectrum_SOI},
with the difference that here the chemical potentials $\mu_\tau$ are opposite at the two edges, $\mu_{1}=-\mu_{\bar 1}$. In this case, the crossed Andreev pairing with amplitude $\Delta_c$ couples a spin-up state at the Fermi level in the upper TI ($\tau=1$, blue arrow) with a spin-down state at the Fermi level in the lower TI 
 ($\tau={\bar 1}$, red arrow), and vice versa.
Again, the crossed pairing is dominant in the regime of strong electron-electron interactions.
 }
\label{spectrum_same_soi}
\end{figure}

As before, the kinetic part of the Hamiltonian is written in terms of slowly moving right- [$R_\tau (x)$] and left- [$L_\tau (x)$] fields defined around the Fermi points $\pm k_F$ as
\begin{align}
H_0 = - i \hbar \upsilon_F \sum_{\tau} \int dx  [R^\dagger_\tau \partial_x R_\tau- L^\dagger_\tau \partial_x L_\tau ].
\end{align}
Here, $ \upsilon_F$ is the Fermi velocity, and the index $\tau$ refers to the corresponding TI. Importantly, the spin up mode corresponds to the right (left) moving field in the upper (lower) sample $\tau=1$ ($\tau=\bar 1$), see Fig. \ref{spectrum_same_soi}. In addition, we choose here the $x$ axis to be codirectional with the propagation direction of the spin down (spin up) mode in the $\tau=1$ ($\tau=\bar 1$) system.

Again, the most important ingredients of our setup are  pairing terms induced  by an ordinary $s$-wave superconductor in proximity contact to the TIs. Such terms are responsible for couplings between states with opposite momenta and spins. We note that the only way to generate a crossed Andreev  term, which pairs modes from different TIs, is to tune the chemical potentials 
to be opposite in the two TIs, $\mu_1=-\mu_{\bar 1}$, see Fig. \ref{spectrum_same_soi}. The corresponding term in the Hamiltonian is written as
\begin{align}
H_c = \int dx \frac{\Delta_c}{2} (R_1^\dagger R^\dagger_{\bar 1} - R^\dagger_{ \bar 1} R_1^\dagger + L_{\bar 1}^\dagger L_{ 1}^\dagger -  L_{ 1}^\dagger L_{\bar 1}^\dagger + H.c.),  \label{hcSOI}
\end{align}
where $\Delta_c$ is the crossed Andreev pairing amplitude. Similarly, we write down the usual pairing term with  amplitude $\Delta_\tau$ induced in each of the TIs separately,
\begin{align}
H_t =\sum_\tau  \int dx\ &\Big[ \frac{\Delta_1}{2} (R_1^\dagger L^\dagger_{ 1} - L^\dagger_{ 1} R_1^\dagger)   \nonumber \\
&+ \frac{\Delta_{\bar 1}}{2} (L_{\bar 1}^\dagger R_{\bar 1}^\dagger -  R_{\bar 1}^\dagger L_{\bar 1}^\dagger)+H.c.\Big]. \label{htauSOI}
\end{align}

For the proposed setup, the crossed Andreev term dominates only along the edge where the two samples are brought close to each other. Along the other edges, $\Delta_c$ is exponentially suppressed at the distance of the coherence length of the superconductor and/or power-law like on the scale of the Fermi wavelength. This results in  interfaces between the  regions dominated by crossed Andreev pairing and the usual pairing regions similar to the ones considered in the previous section, see Fig. \ref{same_soi}.

\subsection{Kramers Pairs of Majorana Fermions}

First we focus on the topological insulators that can be described in an effective single-particle model (i.e. in the absence of explicit interactions).\cite{Hasan_review,Volkov_TI1,Volkov_TI2,Volkov_TI3,Fu_Kane,Zhang_TI} In this case, we can explore the presence of the localized zero-energy  states  by direct diagonalization of the total Hamiltonian $H=H_0+H_c+H_t$ in the fermionic representation. In the basis $\Psi=(R_1, L_1, R_{\bar 1}, L_{\bar 1}, R_1^\dagger, L_1^\dagger, R_{\bar 1}^\dagger, L_{\bar 1}^\dagger)$, the Hamiltonian density is written in terms of the Pauli matrices $\eta_{i}$ ($\tau_{i}$) acting in the electron-hole (channel) space\cite{Braunecker, Klinovaja2012,Klinovaja_Loss_Ladder,Rotating_field} as
\begin{align}
&\mathcal{H}= \hbar \upsilon_F \hat k \rho_3  - \frac{\Delta_{1}}{2} (1+\tau_3) \rho_2 \eta_2 + \frac{\Delta_{\bar 1}}{2} (1-\tau_3) \rho_2 \eta_2 \nonumber \\
&\hspace{110pt} - \Delta_c \eta_2 \tau_2 \rho_3,
\end{align}
where $\rho_{i}$ are the Pauli matrices acting in the left-right mover space.

\begin{figure}[!tb]
\includegraphics[width=0.9\linewidth]{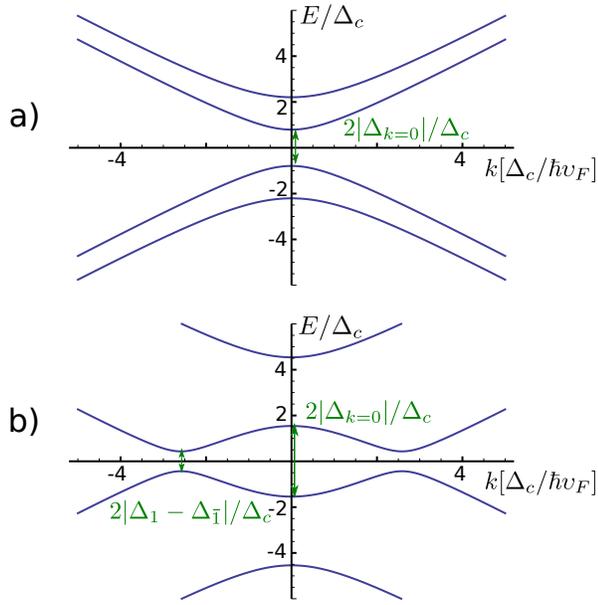}
\caption{The energy spectrum $E(k)$ of the system shown in Fig. \ref{same_soi} around the Fermi points $\pm k_F$ 
[see Eq. (\ref{sp2})]. The gap at $k=0$ closes at $\Delta_c=\sqrt{\Delta_1\Delta_{\bar 1}}$, indicating the topological transition separating (a) the trivial (characterized by $\Delta_{k=0}>0$) and (b) the topological (characterized by $\Delta_{k=0}<0$) phases. We note that (a)  in the trivial phase the gap is minimal at $k=0$, whereas (b)  in the topological phase the gap is minimal at a finite wavevector and is given by $|\Delta_1-\Delta_{\bar 1}|$.
The parameters are chosen to be (a) $\Delta_1/\Delta_c =4$ and $\Delta_{\bar 1}/\Delta_c=2$; (b) $\Delta_1/\Delta_c =2/3$ and $\Delta_{\bar 1}/\Delta_c=1/3$.
}
\label{Gaps}
\end{figure}

The energy spectrum of $\mathcal{H}$ is given by
\begin{widetext}
\begin{align}
E_{\pm,\pm}=  \pm \Big[\Big(\Delta_{1}^2 + \Delta_{\bar 1}^2 +2  \Delta_{c}^2 + 2 (\hbar \upsilon_F k)^2  \pm \sqrt{(\Delta_{1} + \Delta_{\bar 1})^2 [(\Delta_{1} - \Delta_{\bar 1})^2 + 4\Delta_c^2]+ (4 \Delta_c \hbar \upsilon_F k )^2}\Big)/2\Big]^{1/2},  \label{sp2}
\end{align}
\end{widetext}
where each energy level is twofold degenerate, again as a direct consequence of time-reversal invariance of the system, see Fig. \ref{Gaps}.
We note that the system is gapless in two cases: if $\Delta_1=\Delta_{\bar 1}<\Delta_c$ or if $\Delta_c^2=\Delta_1\Delta_{\bar 1}$.  We note here that the spectra of the para- and ortho-helical setups are identical at zero momentum but different at finite momentum [compare Eq. (\ref{spectrum_same_soi}) with Eq. (\ref{sp2})]. Consequently, the gap at $k=0$ defined via Eq. (\ref{gap1}) is again
closed if and only if $\Delta_c^2=\Delta_1\Delta_{\bar 1}$.
In addition to that, in the ortho-helical setup considered in this section, the  gap also closes at finite momenta if $\Delta_1=\Delta_{\bar 1}<\Delta_c$, see Fig.~\ref{Gaps}.  We recall that such finite-momentum closing of the gap does not occur in the para-helical case considered in Sec.~II.

The closing of the gap as a function of the parameters $\Delta_c$ and $\Delta_\tau$ might indicate the presence of a topological phase transition, which we explore further in detail.

In a next step, we consider an interface between two regions: one with finite crossed Andreev pairing term ($\Delta_c>0$ for $x>0$) and one without it ($\Delta_c =0 $ for $x<0$). The usual superconducting terms are assumed be of constant strength $\Delta_\tau$. At the interface $x=0$, there are two zero-energy Majorana bound states $\Psi_{MF1} (x)$ and  $ \Psi_{MF\bar1}(x)$ if $\Delta_c^2>\Delta_1\Delta_{\bar 1}$ and $\Delta_1 \neq \Delta_{\bar 1}$. These two states are time-reversal partners connected by $\Psi_{MF1} (x) = T  \Psi_{MF\bar1}(x)$.
The general form of two MF wavefunctions corresponding to a Kramers pair is written in the basis $\tilde \Psi=(\Psi_{1,\uparrow}, \Psi_{1,\downarrow}, \Psi_{\bar 1,\downarrow}, \Psi_{\bar 1,\uparrow},\Psi_{1,\uparrow}^\dagger, \Psi_{1,\downarrow}^\dagger, \Psi_{\bar 1,\downarrow}^\dagger, \Psi_{\bar 1,\uparrow}^\dagger)$ as
\begin{align}
\Psi_{MF1}(x)=\begin{pmatrix}
f_1(x)\\g_1(x)\\f_{\bar 1}(x)\\g_{\bar 1}(x)\\f_1^*(x)\\g_1^*(x)\\f^*_{\bar 1}(x)\\g^*_{\bar 1}(x)
\end{pmatrix},\ \ \Psi_{MF\bar1}(x)=\begin{pmatrix}
g_1^*(x)\\-f_1^*(x)\\ -g_{\bar 1}^*(x)\\f_{\bar 1}^*(x)
\\g_1(x)\\-f_1(x) \\  -g_{\bar 1}(x)\\f_{\bar 1}(x)
\end{pmatrix},
\end{align}
where $f_{\pm 1}$ and $g_{\pm 1}$ are four arbitrary complex functions (again, we supress the normalization constant).  As before, wavefunctions being rewritten from the basis $\Psi$ to the basis $\tilde \Psi$ just acquire oscillating prefactors $e^{\pm ik_Fx}$.\cite{Klinovaja2012}

In the setup under consideration which assumes a sharp drop of the crossed Andreev pairing strength, these functions are given by
\begin{widetext}
\begin{align}
&f_1(x)=ig_1^*(x)=\begin{cases}
\frac{\Delta_c i e^{ik_F x} ( e^{-x/\xi_1} -e^{-x/\xi_2})}{\sqrt{(\Delta_1 +\Delta_{\bar 1})^2 - 4\Delta_c^2}}, & x>0\\
0,& x<0
                  \end{cases},\\
&f_{\bar1}(x)=ig_{\bar 1}^*(x)=e^{-ik_F x} \times
\begin{cases}
\frac{  e^{-x/\xi_1} (\sqrt{(\Delta_1 +\Delta_{\bar 1})^2 - 4\Delta_c^2}-\Delta_1 -\Delta_{\bar 1})}{2\sqrt{(\Delta_1 +\Delta_{\bar 1})^2 - 4\Delta_c^2}} +\frac{ e^{-x/\xi_2} (\Delta_1 +\Delta_{\bar1}+\sqrt{(\Delta_1 +\Delta_{\bar 1})^2 - 4\Delta_c^2})  }{2\sqrt{(\Delta_1 +\Delta_{\bar 1})^2 - 4\Delta_c^2}},& x>0\\
 e^{x/\xi_3}, & x<0
                               \end{cases}.
\end{align}
\end{widetext}
Here, without loss of generality, we assume that $\Delta_1 >\Delta_{\bar 1}$. The localization lengths $\xi_{1,2,3}$,
\begin{align}
&\xi_1 = 2\hbar \upsilon_F/(\Delta_1-\Delta_{\bar 1}+\sqrt{(\Delta_1+\Delta_{\bar 1})^2-4\Delta_c^2}),\\
&\xi_2 = 2\hbar \upsilon_F/(\Delta_1+\Delta_{\bar 1}+\sqrt{(\Delta_1+\Delta_{\bar 1})^2-4\Delta_c^2}),\\
&\xi_3=\hbar \upsilon_F/\Delta_{\bar1},
\end{align}
are inversely proportional to the gaps in the spectrum opened at the Fermi points $\pm k_F$.\cite{Klinovaja2012}

\subsection{Kramers Pairs of Parafermions in Fractional Topological Insulators}

After having considered the non-interacting model, we focus now on the fractional topological insulators that, again, find their origin in strong electron-electron interactions.\cite{Ady_FTI} Similarly to the previous section, the elementary excitations carry the fractional charge $e/m$, with $m$ a positive odd integer. Again, the fermionic operators $R_\tau$ and $L_\tau$ are redefined in terms of  chiral bosonic  fields $\phi_{r \tau}$, $R_\tau=e^{im\phi_{1\tau}}$ and $L_\tau=e^{im\phi_{\bar1\tau}}$. By analogy, in Nambu representation, we also introduce chiral bosonic operators $\tilde\phi_{1\tau}$ for holes.
The commutation relations between fields are chosen to be
\begin{align}
&[\phi_{r\tau} (x), \phi_{r\tau} (x')]=\frac{i\pi r \tau }{m} {\rm sgn} (x-x'),\label{c1}\\
&[\tilde\phi_{r\tau} (x), \tilde\phi_{r\tau} (x')]=\frac{i\pi r \tau }{m} {\rm sgn} (x-x'),\\
&[\phi_{1 1},\phi_{\bar 1 \bar 1}] =[\phi_{1 1},\tilde\phi_{\bar 1 1}]=[\phi_{\bar 1 \bar 1},\tilde\phi_{1 \bar 1}]=\frac{i\pi }{m},\\
&[\tilde\phi_{1 \bar 1},\phi_{1 1}]=[\tilde\phi_{1 \bar 1},\phi_{1 1}]=[\tilde\phi_{1 \bar 1},\tilde\phi_{\bar 1 1}]=\frac{i\pi }{m},\\
&[\phi_{\bar 1 1},\phi_{ 1 \bar 1}] =[\phi_{\bar 1 1},\tilde\phi_{ 1 1}]=[\phi_{ 1 \bar 1},\tilde\phi_{\bar 1 \bar 1}]=-\frac{i\pi }{m},\\
&[\tilde\phi_{\bar 1 \bar 1},\phi_{\bar 1 1}]=[\tilde\phi_{\bar 1 \bar 1},\phi_{\bar 1 1}]=[\tilde\phi_{\bar 1 \bar 1},\tilde\phi_{ 1 1}]=-\frac{i\pi }{m},
\label{c2}
\end{align}
where $r=\pm 1$.
This choice is made under the constraint that all operators considered in this section that commute with each other in the fermionic representation do also commute in the bosonic representation. In particular, we pay attention to the fact that $H_c$ and $H_\tau$ do not commute with each other whereas $H_1$ commutes with $H_{\bar 1}$.  These observations can be easily checked in the fermionic language [see Eqs. (\ref{hcSOI}) and (\ref{htauSOI})].

Bosonizing $H_\tau$ and $H_c$, we arrive at
\begin{align}
&H_1 = \Delta_1 (\cos [m(\phi_{11}-\tilde\phi_{\bar 11})]-\cos [m(\tilde \phi_{11}-\phi_{\bar 11})]),\\
&H_{\bar 1}=\Delta_{\bar 1} (\cos [m(\tilde \phi_{ 1 \bar 1}-\phi_{\bar 1 \bar 1})]-\cos [m(\phi_{ 1 \bar 1}-\tilde\phi_{\bar 1 \bar 1})]),\\
&H_c=\Delta_c (\cos[m(\phi_{11}-\tilde\phi_{1\bar1})]+\cos[m(\tilde\phi_{\bar 11}-\phi_{\bar 1\bar1})]\nonumber\\
&\hspace{30pt}-\cos[m(\tilde\phi_{11}-\phi_{1\bar1})]-\cos[m(\phi_{\bar 11}-\tilde\phi_{\bar 1\bar1})]).
\end{align}

As already noted in Sec. II, if two operators, for example $H_c$ and $H_\tau$, do not commute with each other, they cannot be ordered simultaneously in the strong coupling regime. Thus, only the coupling constant corresponding to the most relevant term scales up. The competing term scales down and becomes irrelevant.

Further we consider the regime where crossed Andreev pairing dominates in one region ($0<x<L$) while  the usual  pairing dominates in the two neighboring regions ($x<0$ and $x>L$). 
Thus, we obtain the desired interfaces between two non-commuting terms. In addition, we again divide the total Hamiltonian into two parts $H^{(1)}$ and $H^{(\bar 1)}$ that act in two time-reversal symmetry conjugated subspaces: $(\phi_{11},\phi_{\bar 1 \bar 1},\tilde\phi_{\bar 11},\tilde \phi_{ 1 \bar 1})$ and $(\phi_{\bar 11}, \phi_{1\bar 1},\tilde\phi_{11},\tilde\phi_{\bar 1 \bar 1})$.
From now on, we focus only on $H^{(1)}$ and rewrite it in terms of new bosonic fields $\phi_\eta$ and $\theta_\eta$ with $\eta=\pm 1$,
\begin{align}
&\phi_{rr}= \Big[\frac{(\phi_1 + \theta_1)}{m} +  r \phi_{\bar 1} + r \theta_{\bar 1}\Big]/2,\\
&\tilde\phi_{r\bar r}= \Big[\frac{(-\phi_1 + \theta_1)}{m} + r \phi_{\bar 1} - r \theta_{\bar 1}\Big]/2.
\end{align}
As a result, we arrive at
\begin{align}
&H_1^{(1)} = \Delta_1 \cos [\phi_1 + m \phi_{\bar1}],\\
&H_{\bar 1}^{(1)} = \Delta_{\bar 1} \cos [\phi_1 - m \phi_{\bar1}],\\
&H_c^{(1)}=\Delta_c (\cos[\phi_1 +  m\theta_{\bar1}]+\cos[\phi_1 - m\theta_{\bar1}])\nonumber\\
&\hspace{100pt}=2 \Delta_c \cos(\phi_1)\cos(m\theta_{\bar1}).
\end{align}
The commutation relations between the new $\phi$-$\theta$ fields follow from Eqs. (\ref{c1})-(\ref{c2}),
\begin{align}
&[\phi_{\bar 1} (x), \theta_{\bar 1} (x')] = \frac{2i\pi}{m} \theta (x-x'),\\
&[\phi_1,\phi_{\bar 1}]=[\phi_1,\theta_{\bar 1}]=0.
\end{align}
In addition, each field commutes with itself.

To minimize the total energy in the strong coupling regime, the fields standing under the cosines in $H_{\tau}^{(1)}$ and $H_c^{(1)}$ should be uniformly pinned in the corresponding regions where these terms are dominant.\cite{PF_Clarke,PF_Mong} Consequently, the bulk energy spectrum is fully gapped, and the size of the gap is determined by the renormalized value of $\Delta_c$ ($\Delta_1$ and $\Delta_{\bar 1}$) at $0<x<L$ ($x<0$ and $x>L$). The field $\phi_1$ is pinned uniformly, $\phi_1 = \pi \hat M$. The two other fields are pinned inside the corresponding regions,
\begin{align}
&\phi_{\bar 1} =\begin{cases}
                 \pi ( \hat M + 1+2 \hat l_{\bar 1})/m, & x<0\\
                  \pi ( \hat M + 1+2 \hat l_{1})/m, & x>L
                \end{cases},\\
&\theta_{\bar 1} = \pi ( \hat M + 1+2 \hat n)/m,\ \ \ 0<x<L.
\end{align}
Here, $\hat M$, $\hat l_{\pm 1}$, and $\hat n$ are integer valued operators. The only non-zero commutation relation is between $\hat l_1$ and $\hat n$,
\begin{align}
[\hat l_1, \hat n]= \frac{im}{2\pi}. 
\end{align}

Next, we can define two operators $\alpha_{\bar 1}$ acting at $x=0$ and $\alpha_{1}$ acting at $x=L$,
\begin{align}
\alpha_{\bar 1} = e^{i \frac{2\pi}{m}  (\hat l_{\bar 1} + \hat n)}, \ \alpha_{ 1} = - e^{i \frac{2\pi}{m} (\hat l_{1} + \hat n)}.
\end{align}
These two $\alpha$ operators commute with the Hamiltonian and thus correspond to zero-energy states.\cite{PF_Clarke} In addition, they satisfy the commutation relations for ${\mathbb Z}_m$ parafermions,
\begin{align}
&\alpha_1^m=1, \ \alpha_{\bar 1}^m=1,\ \alpha_1 \alpha_{\bar 1} =  \alpha_{\bar 1} \alpha_1 e^{-2\pi i/m}.
\end{align}

To conclude, similarly to the previous section, we identify the ground state degeneracy explicitly. We note that $(\alpha_1^\dagger \alpha_{\bar 1})^m=1$. Thus, the operator $\alpha_1^\dagger \alpha_{\bar 1}$ has $m$ eigenvalues $e^{2\pi i q/m}$ with $q=0, \dots , m-1\  ({\rm mod}\ m)$ corresponding to the eigenstates $\ket q$. This confirms that the ground state of $H_1^{(1)}$ is $m$-fold degenerate,\cite{PF_Clarke,PF_Mong} whereas the ground state of the total time-reversal invariant Hamiltonian $H=H_1^{(1)}+H_{\bar 1}^{(1)}$ consists of $m$ Kramers pairs of localized states.

\section{Conclusions}

We propose two setups in which we generate Kramers partners of Majorana fermions and parafemions. Each setup is based on two topological insulators coupled via proximity to an $s$-wave superconductor. The mutual propagation directions of edge modes is different in the two setups. In the para-helical (ortho-helical) setup, each of the two spin-up modes are moving in the same direction (in opposite directions) along their common edge. Such setups can be realized either by placing two topological insulators with the same spin orbit interaction sign next to each other (ortho-helical setup) or by placing them on  top of each other (para-helical setup).
In addition, the chemical potentials should be adjusted in such a way that crossed Andreev  pairing via the $s$-wave superconductor is made posssible. Moreover, this particular pairing, which is non-local, dominates over the usual one in the regime of strong electron-electron interactions. Taking into account that our scheme relies crucially on crossed Andreev  pairing which has not been demonstrated yet in experiments with topological insulators (but was successfully implemented in experiments on Cooper pair splitters\cite{Schonenberger,Heilblum}), we hope that such tests will be performed in the near future. For examples, signatures of such a pairing can be seen in noise measurements.\cite{Daniel_Yaroslav_PRL_TI_CAS} 
Overall, taking into account that the coupling between topological insulators and superconductors was successfully implemented,\cite{Amir_TI} we believe that inducing both usual and crossed Andreev  pairings is challenging but can be expected to be within experimental reach.

Zero-energy bounds states arise at interfaces between two regions: in one of them the dominant  pairing is the crossed Andreev coupling, while in the other the usual one. As a result, such an interface hosts a Kramers pair of parafermions if the topological insulator supports fractional edge states. Otherwise, it hosts a Kramers pair of Majorana fermions.

Interfaces between regions can be shifted by local gates as it was first proposed for nanowires hosting Majorana fermions.\cite{Alicea_braiding} This allows us to envisage networks based on topological insulator edge modes that can serve as a platform for topological quantum computation. Moreover, the fact that we do not invoke any magnetic fields, which would be detrimental for superconductivity and the topological protection of the edge states in the topological insulators, helps us to extend our setup to a sea of parafermions that could host excitations obeying the Fibonacci type braiding statistics. \cite{PF_Mong,PFs_Loss_2}

\acknowledgments
We thank the UCSB KITP for hospitality where part of this work was performed.
JK acknowledges funding from the
Harvard Quantum Optics Center;
AY  from the NSF through grant
DMR-1206016; a grant from Microsoft
Corporation; and the STC Center for Integrated Quantum
Materials, NSF grant DMR-1231319; and DL
from the Swiss NSF and the NCCR QSIT.


\begin{thebibliography}{100}
\bibitem{Footnote} In spite of the fact that fractional excitations with fractional charge, such as $e/3$, can be constructed explicitly if backscattering terms are taken into account,\cite{kane_PRL,kane,QHE_Klinovaja_PRL,QHE_Klinovaja_PRB} we follow here the approach of Ref. \onlinecite{Ady_FTI} that postulates appropriate commutation relations for  fractional topological insulator  excitations.
\end{thebibliography}
\end{document}